\documentclass[12pt,epsfig]{article}
\usepackage{amsmath}
\usepackage{graphicx}
\usepackage{epstopdf}
\usepackage{cases}
\usepackage{titlesec}
\titleformat*{\section}{\large\bfseries\sffamily}
\titleformat*{\subsection}{\small\bfseries\sffamily}
\begin{document}
	\begin{center}
		Bivariate density estimation using normal-gamma kernel with application to
		astronomy\\
		\vspace{.1 in}
		Uttam Bandyopadhyay$^{1}$ and Soumita Modak$^{2,*}$\\
		\vspace{.05 in}
		Department of Statistics, University of Calcutta\\
		\vspace{.05 in}
		35 Ballygunge Circular Road, Kolkata- 700019, India\\
		\vspace{.05 in}
		email: ubandyopadhyay08@gmail.com$^1$, soumitamodak2013@gmail.com$^2$
	\end{center}
\begin{abstract}
We consider the problem of estimation of a bivariate density function with support $\Re\times[0,\infty)$, where a classical bivariate kernel estimator causes boundary bias due to the non-negative variable. To overcome this problem, we propose four kernel density estimators and compare their performances  in terms of the mean integrated squared error. Simulation study shows that the estimator based on the proposed normal-gamma kernel performs best. Two astronomical data sets are used to demonstrate the applicability of this estimator.
\end{abstract}
\vspace{.1 in}
keywords: {Bivariate density estimation, Product of classical and gamma kernels, $NG$ kernels, Astronomical application.}
\section{Introduction}
Let $\textbf{X}_{i}=(X_{i1}, \ldots, X_{id})^T$, $i=1, \ldots, n$ be $n$ independent realizations of $d$-dimensional random variable $\textbf{X}=(X_{1}, \ldots, X_{d})^T$ having an unknown continuous $d$-variate probability density function $f$. In this chapter, we concentrate on the problem of estimating $f$ by kernel density estimator, in which $f$ with support $\Re^d$ can be estimated by a $d-$variate classical kernel estimator (see, for example, Silverman, 1986; Wand and Jones, 1995). But this causes boundary bias in case of bounded or semi-bounded support. To solve this problem in univariate set-up, the associated kernels are proposed (see, for example, Chen, 1999, 2000; Libengu\'{e}, 2013; Igarashi and Kakizawa, 2014), whereas, in multivariate set-up, the boundary bias can be omitted by using the product of $d$ univariate associated kernels (see, for example, Bouerzmarni and Rombouts, 2010). In the context of multivariate associated kernel, Kokonendji and Som\'{e} (2018) propose a bivariate beta kernel with a correlation structure. Now, when the support is a cartesian product of $\Re$ and bounded or semi-bounded sets, $f$ can be estimated using the product of univariate classical kernels and univariate associated kernels. Here, in particular, we consider the estimation of a bivariate density function with support $\Re\times[0,\infty)$.

In this regard, Section \ref{2} contains the properties of the estimators based on the product of a univariate classical kernel and a univariate gamma kernel. Section \ref{3} provides bivariate density estimators based on normal-gamma ($NG$) kernels. Section \ref{4} discusses the relative performances of the estimators through simulation followed by data study in Section \ref{5}. Section \ref{6} has the discussion, whereas some technical details are deferred to the Appendix.
\section{Product of classical and gamma kernels}\label{2}
Consider a bivariate continuous density function $f$ satisfying (i) $\textbf{x}\in\Re \times [0,\infty)$, (ii) $f$ is twice continuously partially differentiable on $\Re \times [0,\infty)$, (iii) $\int\{\partial f/\partial x_j\}^2 d\textbf{x} < \infty, j=1,2$ and $\int\{x_2\partial^2 f/\partial x^2_2\}^2 d\textbf{x} <\infty $. To estimate $f$, we consider the estimator as follows
\begin{equation}\label{est1}
\hat{f_1}(\textbf{x})=\frac{1}{nh}\sum_{i=1}^{n} K\bigg(\frac{x_1-X_{i1}}{h}\bigg) K_{x_2/b^2+1,b^2}(X_{i2}),
\end{equation}
where $K$ is the classical kernel satisfying (a) $K(-t)=K(t)$, (b) $\int_{-\infty}^{\infty} K(t) dt=1 $, (c) $\int_{-\infty}^{\infty} t K(t) dt=0 $ and (d) $\int_{-\infty}^{\infty} t^2 K(t) dt=k_2\neq 0 $,
with bandwidth $h(>0)$ satisfying $h\rightarrow0$ and $nhb\rightarrow\infty$ as $n\rightarrow\infty$.
$K_{x_2/b^2+1,b^2}$ is the first class of gamma kernels (Chen, $2000$) defined as
\begin{equation*}
K_{x_2/b^2+1,b^2}(t)=\frac{t^{x_2/b^2} e^{-t/b^2}} {b^{2(x_2/b^2+1)} \Gamma (x_2/b^2+1)},
\end{equation*}
where $\Gamma$ is the gamma function,
with bandwidth $b^2(>0)$ satisfying $b\rightarrow0$ and $nhb\rightarrow\infty$ as $n\rightarrow\infty$. Bandwidths of the kernels are so chosen as to make the amount of smoothing in the same scale for both the kernels. In general, any associated kernel can be used here. However, we choose the gamma kernel due to its flexible properties (see, for example, Chen, $2000$).

Now, using $y_1=x_1-ht$, we get
\begin{align}\label{Expc}
E\{\hat{f_1}(\textbf{x})\}&=\int_{-\infty}^{\infty} \int_{0}^{\infty}\bigg[ h^{-1}K\bigg(\frac{x_{1}-y_{1}}{h}\bigg)\bigg] K_{x_2/b^2+1,b^2}(y_{2}) f(y_{1},y_{2}) dy_{1} dy_{2} \nonumber\\
&=\int_{-\infty}^{\infty} \int_{0}^{\infty} K(t) K_{x_2/b^2+1,b^2}(y_{2}) f(x_1-h t,y_2) dt dy_2 \nonumber\\
&=\int_{-\infty}^{\infty} K(t) E_{\xi_{x_2}}[f(x_1-ht,y_2)] dt,
\end{align}
where $\xi_{x_2}$ follows gamma$(x_{2}/b^{2}+1,b^2)$.

Again, Taylor series expansion gives $E_{\xi_{x_2}}[f(x_1-ht,\xi_{x_2})]$ as
\begin{align*}
\lefteqn{\hspace{.165in}f(x_1,x_2)+ E_{\xi_{x_2}}(x_1-ht-x_1) f^1(\textbf{x})
	+E_{\xi_{x_2}}(\xi_{x_2}-x_2) f^2(\textbf{x})}\\
&\hspace{.165in}+\frac{1}{2}E_{\xi_{x_2}}\{(x_1-ht-x_1)^2\} f^{11}(\textbf{x})+\frac{1}{2}E_{\xi_{x_2}}\{(\xi_{x_2}-x_2)^2\} f^{22}(\textbf{x})\\
&\hspace{.165in}+\frac{1}{2}E_{\xi_{x_2}}\{(-ht)(\xi_{x_2}-x_2)\} f^{12}(\textbf{x}) +\frac{1}{2}E_{\xi_{x_2}}\{(-ht)(\xi_{x_2}-x_2)\} f^{21}(\textbf{x})\\
&\hspace{.165in}+o(h^2+b^2)\\ 
&=f(\textbf{x})-htf^1(\textbf{x})+ E_{\xi_{x_2}}(\xi_{x_2}-x_2-b^2) f^2(\textbf{x})+b^2f^2(\textbf{x})\\
&\hspace{.165in}+\frac{1}{2}h^2t^2f^{11}(\textbf{x})+\frac{1}{2}Var(\xi_{x_2})f^{22}(\textbf{x})+o(h^2+b^2),
\end{align*}
where $f^j=\partial f/ \partial x_j$ and $f^{jj^{\prime}}=\partial^2 f/ \partial x_{j^{\prime}} \partial x_j , j,j^{\prime}=1, 2$. Then,
substituting the last expression in \eqref{Expc}, we get
\begin{equation*}
E\{\hat{f_1}(\textbf{x})\}=f(\textbf{x})+\frac{1}{2}k_2h^2f^{11}(\textbf{x})+b^2 \{f^2(\textbf{x})+\frac{1}{2}x_2 f^{22}(\textbf{x})\}+o(h^2+b^2),
\end{equation*}
which implies $Bias\{\hat{f_1}(\textbf{x})\}$ is
\begin{equation}\label{B}
\frac{1}{2}k_2h^2f^{11}(\textbf{x})+b^2 \{f^2(\textbf{x})+\frac{1}{2}x_2 f^{22}(\textbf{x})\}+o(h^2+b^2)=O(h^2+b^2).
\end{equation}
This shows estimator $\hat{f_1}$ is free of boundary bias and the corresponding integrated squared bias is given by
\begin{align}\label{ISB}
	\lefteqn{\int\{Bias(\hat{f_1}(\textbf{x}))\}^2 d\textbf{x}}\nonumber\\
	&=\frac{1}{4}k_2^2h^4 \int\{f^{11}(\textbf{x})\}^2d\textbf{x}+b^4 \int\{f^2(\textbf{x})+\frac{1}{2}x_2 f^{22}(\textbf{x})\}^2 d\textbf{x} \nonumber\\
&\hspace{.165in}+k_2h^2b^2\int f^{11}(\textbf{x})\{f^2(\textbf{x})+\frac{1}{2}x_2 f^{22}(\textbf{x})\}d\textbf{x}+o(h^4+b^4).
\end{align}

Now,
\begin{align*}
Var\{\hat{f_1}(\textbf{x})\}&=n^{-1} Var\{K(X_{i1})K_{x_2/b^2+1,b^2}(X_{i2})\}\\
&=n^{-1}E\{ K(X_{i1})K_{x_2/b^2+1,b^2}(X_{i2})\}^2+O(n^{-1})\\
&=n^{-1}h^{-1} \int_{-\infty}^{\infty} \int_{0}^{\infty} K^2(t) K^2_{x_2/b^2+1,b^2}(y_{2}) f(x_1-ht,y_2) dt dy_2\\
&\hspace{.165in}+O(n^{-1})
\end{align*}
and
\begin{align*}
\int_{0}^{\infty} K^2_{x_2/b^2+1,b^2}(y_{2}) f(x_1-ht,y_2) dy_2
=B_b(x_2) E_{\eta_{x_2}}\{f(x_1-ht,\eta_{x_2})\},
\end{align*}
where $\eta_{x_2}$ follows gamma$(2x_2/b^2+1,b^2)$ and $B_b(x_2)=\frac{b^{-2}\Gamma(2x_2/b^2+1)}{2^{2x_2/b^2+1}\Gamma^2(x_2/b^2+1)}$.
Lemma $3$ of Brown and Chen $(1999)$ gives
\begin{numcases}
{B_b(x_2)\sim}\nonumber
\frac{1}{2\sqrt{\pi}}  b^{-1} x^{-1/2} & if  $\frac{x_2}{b^2}\rightarrow\infty$,\\ \nonumber
\frac{\Gamma(2\kappa+1)}{2^{1+2\kappa} \Gamma^2(\kappa+1)}  b^{-2} & if $\frac{x_2}{b^2}\rightarrow \kappa$ (a non-negative constant),
\end{numcases}
which implies
\begin{numcases}
{Var\{\hat{f_1}(\textbf{x})\}\sim}\nonumber\label{Var}
\frac{1}{2\sqrt{\pi}} n^{-1}h^{-1} b^{-1}k_3 x^{-1/2}_2 f(\textbf{x}) & if $\frac{x_2}{b^2}\rightarrow\infty$,\\ \nonumber
\frac{ \Gamma(2\kappa+1)}{2^{1+2\kappa} \Gamma^2(\kappa+1)} n^{-1}h^{-1} b^{-2}k_3 f(\textbf{x}) & if $\frac{x_2}{b^2}\rightarrow \kappa$,\\
\end{numcases}
where $k_3=\int_{-\infty}^{\infty}  K^2(t) dt$. Expressions \eqref{B} and \eqref{Var} imply that for $h\rightarrow0, b\rightarrow0,$ and $ nhb\rightarrow\infty$ as $n\rightarrow\infty$, the nonparametric density estimator $\hat{f_1}(\textbf{x})$ is consistent for the true density function $f(\textbf{x})$ at each point $\textbf{x}$. Now, for $\delta=b^{2-\epsilon}$ with $1<\epsilon<2$, 
\begin{align}
\lefteqn{\int_{-\infty}^{\infty} \int_{0}^{\infty} Var\{\hat{f_1}(\textbf{x})\} dx_1 dx_2}\nonumber\\
& =  \int_{-\infty}^{\infty} \int_{0}^{\delta} Var\{\hat{f_1}(\textbf{x})\} dx_1 dx_2+\int_{-\infty}^{\infty} \int_{\delta}^{\infty} Var\{\hat{f_1}(\textbf{x})\} dx_1 dx_2\nonumber \\
& =  \frac{1}{2\sqrt{\pi}} n^{-1}h^{-1} b^{-1}k_3 \int_{-\infty}^{\infty}  \int_{\delta}^{\infty}x^{-1/2}_2 f(\textbf{x}) dx_1 dx_2+O(n^{-1}h^{-1}b^{-\epsilon}) \nonumber\\
&=\frac{1}{2\sqrt{\pi}} n^{-1}h^{-1} b^{-1}k_3 \int_{-\infty}^{\infty}  \int_{0}^{\infty}x^{-1/2}_2 f(\textbf{x}) dx_1 dx_2+ o(n^{-1}h^{-1}b^{-1}),
\label{IV}
\end{align}
provided $\int_{-\infty}^{\infty}  \int_{0}^{\infty}x^{-1/2}_2 f(\textbf{x}) dx_1 dx_2$ is finite.

Combining \eqref{ISB} and \eqref{IV}, the mean integrated squared error (MISE) is obtained as
\begin{align}
\lefteqn{MISE\{\hat{f_1}(\textbf{x})\}=\int\{Bias(\hat{f_1}(\textbf{x}))\}^2 d\textbf{x}+\int Var\{\hat{f_1}(\textbf{x})\} d\textbf{x}} \nonumber\\
&=\frac{1}{4}k_2^2h^4 \int\{f^{11}(\textbf{x})\}^2d\textbf{x}+ b^4 \int\{f^2(\textbf{x})+\frac{1}{2}x_2 f^{22}(\textbf{x})\}^2 d\textbf{x} \nonumber\\
&\hspace{.165in}+ k_2h^2b^2\int f^{11}(\textbf{x})\{f^2(\textbf{x})+\frac{1}{2}x_2 f^{22}(\textbf{x})\}d\textbf{x} \nonumber\\
&\hspace{.165in} +\frac{1}{2\sqrt{\pi}} n^{-1}h^{-1} b^{-1}k_3\int  x^{-1/2}_2 f(\textbf{x})d\textbf{x}+ o(n^{-1}h^{-1}b^{-1}+h^4+b^4)
\label{mise}
\end{align}
and the leading terms in (\ref{mise}) give the expression of the corresponding asymptotic mean integrated squared error (AMISE).
AMISE is optimal for $h_{opt}=h_0 n^{-1/6}$ and $b_{opt}=b_0 n^{-1/6}$, where $h_0$ and $b_0$ are constants, i.e. the optimal bandwidths for kernel density estimator $\hat{f_1}$ are $O(n^{-1/6})$ and $O(n^{-1/3})$ which give $AMISE\{\hat{f_1}(\textbf{x})\}_{opt}$ as
\begin{align*}
\lefteqn{\bigg[\frac{1}{4}k_2^2h_0^4 \int\{f^{11}(\textbf{x})\}^2d\textbf{x}
+ b_0^4 \int\{f^2(\textbf{x})+\frac{1}{2}x_2 f^{22}(\textbf{x})\}^2 d\textbf{x}}\\ &\hspace{.165in}+ k_2h_0^2b_0^2\int f^{11}(\textbf{x})\{f^2(\textbf{x})+\frac{1}{2}x_2 f^{22}(\textbf{x})\}d\textbf{x}\\
&\hspace{.165in}+\frac{1}{2\sqrt{\pi}} n^{-1}h_0^{-1} b_0^{-1}k_3\int  x^{-1/2}_2 f(\textbf{x})d\textbf{x}\bigg] n^{-2/3}.
\end{align*}
For $h=b=h_0$, the optimal $h_0$ is 
\begin{align*}
\Bigg[\frac{\frac{1}{2\sqrt{\pi}}k_3\int  x^{-1/2}_2 f(\textbf{x})d\textbf{x}}{2\int\{\frac{1}{2}k_2f^{11}(\textbf{x})+f^2(\textbf{x})+\frac{1}{2}x_2f^{22}(\textbf{x})\}^2d\textbf{x}}\Bigg]^{1/6}n^{-1/6},
\end{align*}
which gives $AMISE_{opt}$ as
\begin{align*}
\lefteqn{\frac{3}{2^{2/3}}\bigg[\int\bigg\{\frac{1}{2}k_2f^{11}(\textbf{x})+f^2(\textbf{x})+\frac{1}{2}x_2f^{22}(\textbf{x})\bigg\}^2d\textbf{x}\bigg]^{1/3}}\\ &\bigg[\frac{1}{2\sqrt{\pi}}k_3\int  x^{-1/2}_2 f(\textbf{x})d\textbf{x}\bigg]^{2/3} n^{-2/3}.
\end{align*}
\label{B12}

Another estimator of $f$ is considered as
\begin{equation}\label{est12}
\hat{f_2}(\textbf{x})=\frac{1}{nh}\sum_{i=1}^{n} K(\frac{x_1-X_{i1}}{h}) K_{\rho_{b^2}(x_2),b^2}(X_{i2}),
\end{equation}
where $K_{\rho_{b^2}(x),b^2}$ is the second class of gamma kernels (Chen, $2000$) defined as
\begin{numcases}
{\rho_{b^2}(x_2) =}
x_2/b^2       & if  $x_2\geq2b^2$, \nonumber\\
\frac{1}{4}(x_2/b^2)^2+1  & if $x_2\in[0,2b^2)$.
\end{numcases}
So, $Bias\{\hat{f_2}(\textbf{x})\}$, given by
\begin{numcases}
{} 
\frac{1}{2}k_2h^2f^{11}(\textbf{x})+\frac{1}{2}b^2 x_2f^{22}(\textbf{x})+o(h^2+b^2) & if $x_2\geq2b^2$, \nonumber\\
\frac{1}{2}k_2h^2f^{11}(\textbf{x})+b^2 \{\rho_{b^2}(x_2)-x_2/b^2\}f^2(\textbf{x}) & if $x_2\in[0,2b^2)$, \nonumber\\
+o(h^2+b^2) 
\end{numcases}\label{B12}
which shows the boundary unbiasedness of estimator $\hat{f_2}$ and for a non-negative constant $\kappa$,
\begin{numcases}
{Var\{\hat{f_2}(\textbf{x})\}\sim} \nonumber
\frac{1}{2\sqrt{\pi}} n^{-1}h^{-1} b^{-1}k_3 x^{-1/2}_2 f(\textbf{x}) & if $\frac{x_2}{b^2}\rightarrow\infty$,\\ \nonumber
\frac{\Gamma(\kappa^2/2+1)}{2^{1+\kappa^2/2} \Gamma^2(\kappa^2/4+1)} n^{-1}h^{-1} b^{-2}k_3 f(\textbf{x}) & if $\frac{x_2}{b^2}\rightarrow \kappa$,
\end{numcases}
imply
\begin{align*}
MISE\{\hat{f_2}(\textbf{x})\}
&=\frac{1}{4}k_2^2h^4 \int\{f^{11}(\textbf{x})\}^2d\textbf{x}+\frac{1}{4}b^4 \int\{x_2f^{22}(\textbf{x})\}^2 d\textbf{x}\\
&\hspace{.165in}+ \frac{1}{2}k_2h^2b^2\int f^{11}(\textbf{x})\{x_2f^{22}(\textbf{x})\}d\textbf{x}+o(h^4+b^4)\\
&\hspace{.165in}+\frac{1}{2\sqrt{\pi}} n^{-1}h^{-1} b^{-1}k_3\int  x^{-1/2}_2 f(\textbf{x})d\textbf{x}+o(n^{-1}h^{-1}b^{-1}).
\end{align*}
For $h=b=h_0$, the optimal $h_0$ is 
\begin{align*}
\Bigg[\frac{\frac{1}{2\sqrt{\pi}}k_3\int  x^{-1/2}_2 f(\textbf{x})d\textbf{x}}{\frac{1}{2}\int\{k_2f^{11}(\textbf{x})+ x_2f^{22}(\textbf{x})\}^2d\textbf{x}}\Bigg]^{1/6}n^{-1/6},
\end{align*}
which corresponds to $AMISE_{opt}$, given by
\begin{align*}
\frac{3}{2^{4/3}}\bigg[\int\bigg\{k_2f^{11}(\textbf{x})+ x_2f^{22}(\textbf{x})\bigg\}^2d\textbf{x}\bigg]^{1/3} \bigg[\frac{1}{2\sqrt{\pi}}k_3\int  x^{-1/2}_2 f(\textbf{x})d\textbf{x}\bigg]^{2/3} n^{-2/3}.
\end{align*}
Observe that $AMISE_{opt}\{\hat{f_1}(\textbf{x})\} \geq AMISE_{opt}\{\hat{f_2}(\textbf{x})\}$ implies $\hat{f_2}$ is expected to have a better asymptotic performance than $\hat{f_1}$.
\section{Bivariate density estimation using $NG$ kernel}\label{3}
Consider the density function $K_{\Theta}$ of a bivariate normal-gamma distribution defined as (Bernardo and Smith, $2000$)
\begin{align}\label{NgK}
K_{\Theta}(t_1,t_2)&= NG(t_1,t_2|\Theta=(\mu, \lambda, \alpha, \beta))=N(t_1|\mu,(\lambda t_2)^{-1}) Ga(t_2|\alpha,\beta) \nonumber\\
&=\sqrt{\frac{\lambda t_2}{2\pi }}e^{-\frac{(t_1-\mu)^2\lambda t_2}{2}}\times \frac{\beta^{\alpha}}{\Gamma(\alpha)} t_{2}^{\alpha-1} e^{-\beta t_2}
\end{align}
with $\mu\in\Re, \lambda>0, \alpha>0, \beta>0$, where $N$ and $Ga$, respectively, stand for normal and gamma distributions. Using \eqref{NgK}, we define the following estimator of $f$ as
\begin{equation}\label{est2}
\hat{f_3}(\textbf{x})=\frac{1}{n}\sum_{i=1}^{n} K_{\Theta_1}(\textbf{X}_{i}),
\end{equation}
where $K_{\Theta_1}$ is the $NG$ kernel with $\Theta=\Theta_1=(x_1, 1/(|x_1|b_1+b_1^2)(x_2+b_2), x_2/b_2+2, 1/b_2)$ such that the bandwidths $b_1,b_2\rightarrow0$ and $nb_1b_2\rightarrow\infty$ as $n\rightarrow\infty.$ Then,
\begin{equation*}
E\{\hat{f_3}(\textbf{x})\}=\int K_{\Theta_1}(\textbf{y}) d\textbf{y}=E\{f(\boldsymbol{\xi}_x)\},
\end{equation*}
where $\boldsymbol{\xi}_x=(\xi_{x_1},\xi_{x_2})^T$ follows $NG(\xi_{x_1},\xi_{x_2}|x_1, 1/(|x_1|b_1+b_1^2)(x_2+b_2), x_2/b_2+2,1/b_2)$,
which implies $E(\xi_{x_1})=x_1, E(\xi_{x_1})=x_2+2b_2, Var(\xi_{x_1})=|x_1|b_1+b_1^2$ and $Var(\xi_{x_2})=x_2b_2+2b_2^2$.
By Taylor series expansion we get (see, Appendix A.1),
\begin{align*}
E\{f(\boldsymbol{\xi}_x)\}&=f(\textbf{x})+E(\xi_{x_1}-x_1)f^1(\textbf{x})+E(\xi_{x_2}-x_2)f^2(\textbf{x})\\
&\hspace{.165in}+\frac{1}{2}E\{(\xi_{x_1}-x_1)(\xi_{x_2}-x_2)\} \{f^{12}(\textbf{x})+f^{21}(\textbf{x})\}\\
&\hspace{.165in}+\frac{1}{2}E(\xi_{x_1}-x_1)^2f^{11}(\textbf{x})+\frac{1}{2}E(\xi_{x_2}-x_2)^2f^{22}(\textbf{x})+o(b_1+b_2)
\\ \nonumber
&=f(\textbf{x})+b_1\{\frac{1}{2}|x_1|f^{11}(\textbf{x})\}+b_2\{2f^2(\textbf{x})+\frac{1}{2}x_2f_{22}(\textbf{x})\}+o(b_1+b_2).
\end{align*}
Therefore, $Bias\{\hat{f_3}(\textbf{x})\}$ is given by
\begin{equation}\label{B2}
b_1\{\frac{1}{2}|x_1|f^{11}(\textbf{x})\}+b_2\{2f^2(\textbf{x})+\frac{1}{2}x_2f^{22}(\textbf{x})\}+o(b_1+b_2)=O(b_1+b_2),
\end{equation}
which shows estimator $\hat{f}_3$ is free of boundary bias, and the integrated squared bias is
\begin{align}\label{ISB2}
\lefteqn{\int\{Bias(\hat{f_3}(\textbf{x}))\}^2 d\textbf{x}}\nonumber\\
&=\frac{1}{4}b_1^2\int\{x_1f^{11}(\textbf{x})\}^2d\textbf{x} + b_2^2 \int\{2f^2(\textbf{x})+\frac{1}{2}x_2 f^{22}(\textbf{x})\}^2 d\textbf{x} \nonumber\\
&\hspace{.165in}+ b_1b_2\int\{|x_1|f^{11}(\textbf{x})\}\{2f^2(\textbf{x})+\frac{1}{2}x_2f_{22}(\textbf{x})\}d\textbf{x}+o(b_1^2+b_2^2).
\end{align}

The variance of $\hat{f_3}(\textbf{x})$ is
\begin{tiny}
	\begin{numcases}
	{Var\{\hat{f_3}(\textbf{x})\} \sim} \nonumber
	\frac{1}{4\pi\sqrt{ e}}n^{-1}b_1^{-1/2}  b_2^{-1/2}|x_1|^{-1/2} x_2^{-1/2} f(\textbf{x}) &if $|x_1|/b_1\rightarrow\infty$, $x_2/b_2\rightarrow\infty,$\\ \nonumber
	\frac{\Gamma(2\kappa_2+7/2)}{\sqrt{\pi}2^{2\kappa_2+9/2}\sqrt{(\kappa_1+1)(\kappa_2+1)}\Gamma^2(\kappa_2+2)} n^{-1}b_1^{-1} b_2^{-1}  f(\textbf{x}) &if $|x_1|/b_1\rightarrow\kappa_1$, $x_2/b_2\rightarrow\kappa_2,$\\ \nonumber
	\frac{\Gamma(2\kappa_2+7/2)}{\sqrt{\pi}2^{2\kappa_2+9/2}\sqrt{(\kappa_2+1)}\Gamma^2(\kappa_2+2)}n^{-1} b_1^{-1/2} b_2^{-1}|x_1|^{-1/2}  f(\textbf{x}) &if $|x_1|/b_1\rightarrow\infty$, $x_2/b_2\rightarrow\kappa_2,$\\ \nonumber
	\frac{1}{4\pi\sqrt{ e}\sqrt{\kappa_1+1}}n^{-1}b_1^{-1}  b_2^{-1/2} x_2^{-1/2} f(\textbf{x}) &if $|x_1|/b_1\rightarrow\kappa_1$, $x_2/b_2\rightarrow\infty$,
	\end{numcases}
\end{tiny}
for non-negative constants $\kappa_1,\kappa_2$ (see, Appendix A.2), and
\begin{align}\label{IV2}
\lefteqn{\int Var\{\hat{f_3}(\textbf{x})\}d\textbf{x}}\nonumber\\
&=\frac{1}{4\pi\sqrt{ e}}n^{-1}b_1^{-1/2}  b_2^{-1/2}\int|x_1|^{-1/2} x_2^{-1/2} f(\textbf{x}) d\textbf{x}+o(n^{-1}b_1^{-1/2}b_2^{-1/2}),
\end{align}
assuming $\int|x_1|^{-1/2} x_2^{-1/2} f(\textbf{x}) d\textbf{x}<\infty$ (see, Appendix A.3).

Now, combining (\ref{ISB2}) and (\ref{IV2}), we get the expression of the MISE as follows
\begin{align*}
MISE\{\hat{f_3}(\textbf{x})\}&= \frac{1}{4}b_1^2\int\{|x_1|f^{11}(\textbf{x})\}^2d\textbf{x}+ b_2^2 \int\{2f^2(\textbf{x})+\frac{1}{2}x_2 f^{22}(\textbf{x})\}^2 d\textbf{x}  \\
&\hspace{.165in}+b_1b_2\int|x_1|f^{11}(\textbf{x})\{2f^2(\textbf{x})+\frac{1}{2}x_2f_{22}(\textbf{x})\}d\textbf{x} \\
&\hspace{.165in}+\frac{1}{4\pi\sqrt{ e}}n^{-1}b_1^{-1/2}  b_2^{-1/2}\int|x_1|^{-1/2} x_2^{-1/2} f(\textbf{x}) d\textbf{x}\\
&\hspace{.165in}+o(b_1^2+b_2^2+n^{-1}b_1^{-1/2}b_2^{-1/2}).\\
\end{align*}
For $b_1=b_2=b_0$, the optimal $b_0$ is given by
\begin{align*}
\bigg[\frac{\frac{1}{4\pi\sqrt{ e}}\int|x_1|^{-1/2} x_2^{-1/2} f(\textbf{x}) d\textbf{x}}{2\int\{\frac{1}{2}|x_1|f^{11}(\textbf{x})+2f^2(\textbf{x})+\frac{1}{2}x_2f^{22}(\textbf{x})\}^2d\textbf{x}}\bigg]^{1/3}n^{-1/3},
\end{align*}
for which the optimal AMISE is obtained as
\begin{align*}
AMISE_{opt}&=\frac{3}{2^{2/3}}\bigg[\int \bigg\{\frac{1}{2}|x_1|f^{11}(\textbf{x})+2f^2(\textbf{x})+\frac{1}{2}x_2f^{22}(\textbf{x})\bigg\}^2d\textbf{x}\bigg]^{1/3}\\
&\hspace{.165in}\bigg[\frac{1}{4\pi\sqrt{ e}}\int|x_1|^{-1/2} x_2^{-1/2} f(\textbf{x}) d\textbf{x}\bigg]^{2/3}n^{-2/3}.
\end{align*}

We propose another estimator of $f$ using \eqref{NgK} as follows
\begin{equation}\label{est22}
\hat{f_4}(\textbf{x})=\frac{1}{n}\sum_{i=1}^{n} K_{\Theta_2}(\textbf{X}_{i}),
\end{equation}
where  $K_{\Theta_2}$ is the $NG$ kernel with $\Theta=\Theta_2=(x_1, 1/(|x_1|b_1+b_1^2)(x_2+b_2), \alpha_{b_2}(x_2), 1/b_2)$ and
\[ \alpha_{b_2}(x_2) =
\begin{cases}
x_2/b_2                   & \quad \text{if } x_2\geq3b_2,\\
\frac{1}{9}(x_2/b_2)^2+2  & \quad \text{if } x_2\in[0,3b_2),\\
\end{cases}
\]
such that $b_1,b_2\rightarrow0,nb_1b_2\rightarrow\infty$ as $n\rightarrow\infty.$
Then, in the similar fashion as in estimator $\hat{f_3}$, we obtain $Bias\{\hat{f_4}(\textbf{x})\}$ as
\[ 
\begin{cases}
\frac{1}{2}\{b_1|x_1|f^{11}(\textbf{x})+b_2x_2f^{22}(\textbf{x})\}+o(b_1+b_2)                              & \quad \text{if } x_2\geq3b_2,\\
\frac{1}{2}b_1|x_1|f^{11}(\textbf{x})+b_2\{\alpha_{b_2}(x_2)-\frac{x_2}{b_2}\}f^2(\textbf{x})+o(b_1+b_2)   & \quad \text{if } x_2\in[0,3b_2),\\
\end{cases}
\]
establishing its boundary unbiasedness, and
\begin{tiny}
	\[ Var\{\hat{f_4}(\textbf{x})\} \sim
	\begin{cases}
	\frac{1}{4\pi\sqrt{ e}}n^{-1}b_1^{-1/2}  b_2^{-1/2}|x_1|^{-1/2} x_2^{-1/2} f(\textbf{x})         & \quad \text{if } |x_1|/b_1\rightarrow\infty, \quad x_2/b_2\rightarrow\infty,\\
	\frac{\Gamma(2/9\kappa_2^2+7/2)}{\sqrt{\pi}2^{2/9\kappa_2^2+9/2}\sqrt{(\kappa_1+1)(\kappa_2+1)}\Gamma^2(1/9\kappa_2^2+2)} n^{-1}b_1^{-1} b_2^{-1}  f(\textbf{x})    & \quad \text{if }  |x_1|/b_1\rightarrow\kappa_1, \quad x_2/b_2\rightarrow\kappa_2,\\
	\frac{\Gamma(2/9\kappa_2^2+7/2)}{\sqrt{\pi}2^{2/9\kappa_2^2+9/2}\sqrt{(\kappa_2+1)}\Gamma^2(1/9\kappa_2^2+2)}n^{-1} b_1^{-1/2} b_2^{-1}|x_1|^{-1/2}  f(\textbf{x})    & \quad \text{if }  |x_1|/b_1\rightarrow\infty, \quad x_2/b_2\rightarrow\kappa_2,\\
	\frac{1}{4\pi\sqrt{ e}\sqrt{\kappa_1+1}}n^{-1}b_1^{-1}  b_2^{-1/2} x_2^{-1/2} f(\textbf{x})       & \quad \text{if } |x_1|/b_1\rightarrow\kappa_1, \quad x_2/b_2\rightarrow\infty,\\
	\end{cases}
	\]
\end{tiny}
for non-negative constants $\kappa_1$, $\kappa_2$. Hence, 
\begin{align*}
MISE\{\hat{f_4}(\textbf{x})\}&=\frac{1}{4}b_1^2\int\{x_1f^{11}(\textbf{x})\}^2d\textbf{x}+ \frac{1}{4}b_2^2 \int\{x_2f^{22}(\textbf{x})\}^2 d\textbf{x} \\
	&\hspace{.165in}+\frac{1}{2}b_1b_2\int|x_1|x_2f^{11}(\textbf{x})f^{22}(\textbf{x})d\textbf{x}
+o(b_1^2+b_2^2)\\
&\hspace{.165in}+\frac{1}{4\pi\sqrt{ e}}n^{-1}b_1^{-1/2}  b_2^{-1/2}\int|x_1|^{-1/2} x_2^{-1/2} f(\textbf{x}) d\textbf{x}\\
&\hspace{.165in}+o(n^{-1}b_1^{-1/2}b_2^{-1/2}).
\end{align*}
For $b_1=b_2=b_0$, the optimal $b_0$ is given by
\begin{align*}
\bigg[\frac{\frac{1}{4\pi\sqrt{ e}}\int|x_1|^{-1/2} x_2^{-1/2} f(\textbf{x}) d\textbf{x}}{\frac{1}{2}\int\{|x_1|f^{11}(\textbf{x})+x_2f^{22}(\textbf{x})\}^2d\textbf{x}}\bigg]^{1/3}n^{-1/3}
\end{align*}
and, therefore, the optimal AMISE is obtained as
\begin{align*}
AMISE_{opt}&=\frac{3}{2^{4/3}}\bigg[\int \bigg\{|x_1|f^{11}(\textbf{x})+x_2f^{22}(\textbf{x})\bigg\}^2d\textbf{x}\bigg]^{1/3}\\
&\hspace{.165in}\bigg[\frac{1}{4\pi\sqrt{ e}}\int|x_1|^{-1/2} x_2^{-1/2} f(\textbf{x}) d\textbf{x}\bigg]^{2/3}n^{-2/3}.
\end{align*}
\section{Simulation study}\label{4}
\begin{figure}
		\centering
		\includegraphics[width=1\textwidth]{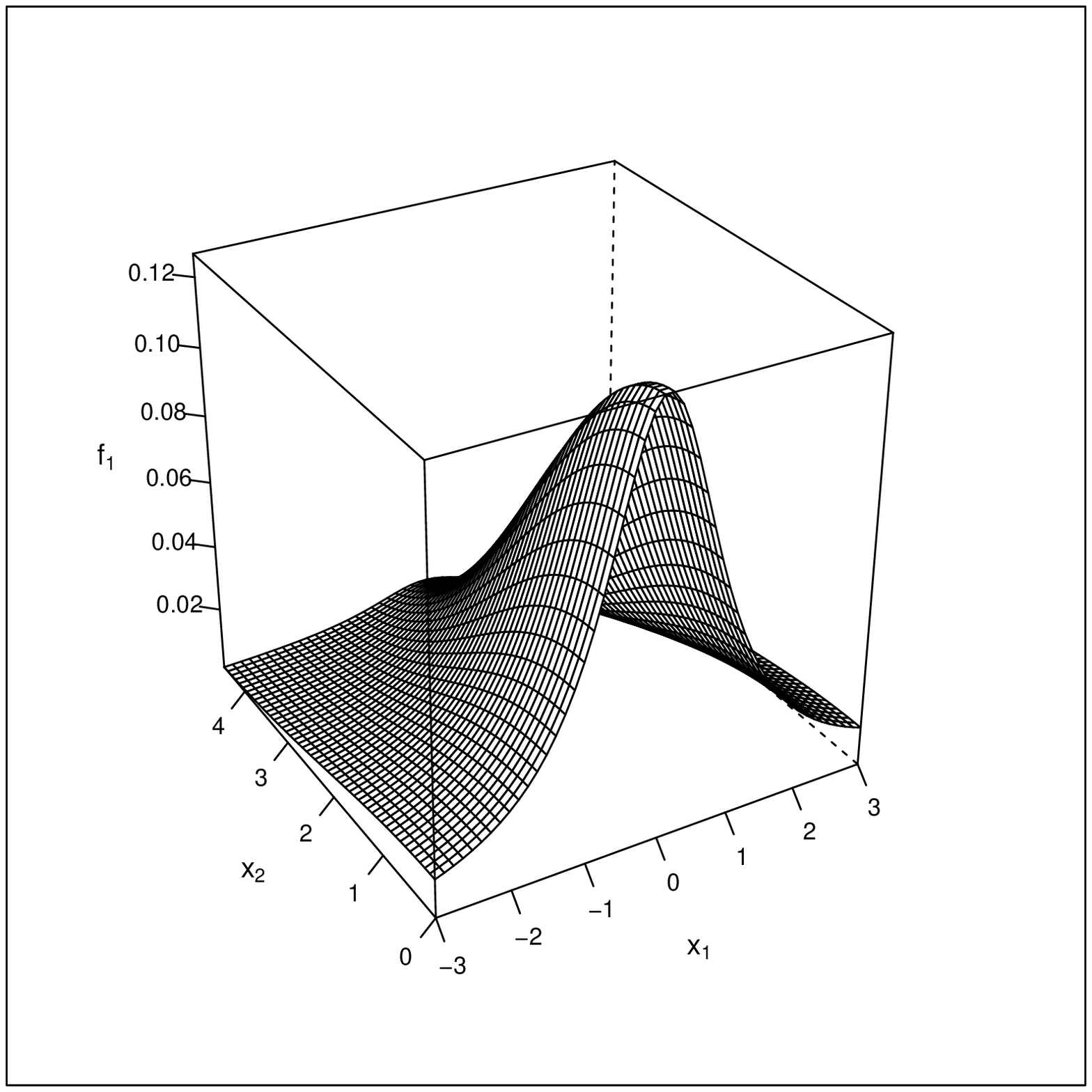}
		\caption{Surface plot of the distribution $f_1$.}\label{f1}
	\end{figure}

In simulation study, we consider $n=100$ and $200$ with $1,000$ replications for the following target distributions of different shapes, where $C, HN, N, Ga$, $Exp, LG, TN$ represent Cauchy, Half-Normal, Normal, Gamma, Exponential, Logistic and Truncated
Normal (truncated at zero) distributions respectively. Here $l, s, sh$ and $rt$ respectively denote the location, scale, shape and rate of the corresponding distribution. \\
	$1)$\hspace{.1 in}Product of $C (l=0, s=1)$ and $HN (s=2)$, say, $f_{1}$ (Fig. \ref{f1}). \\
	$2)$\hspace{.1 in}Product of $0.2 \times N (l=-3, s=1)+ 0.6 \times C (l=0, s=1)$
	
\begin{figure}
		\centering
		\includegraphics[width=1\textwidth]{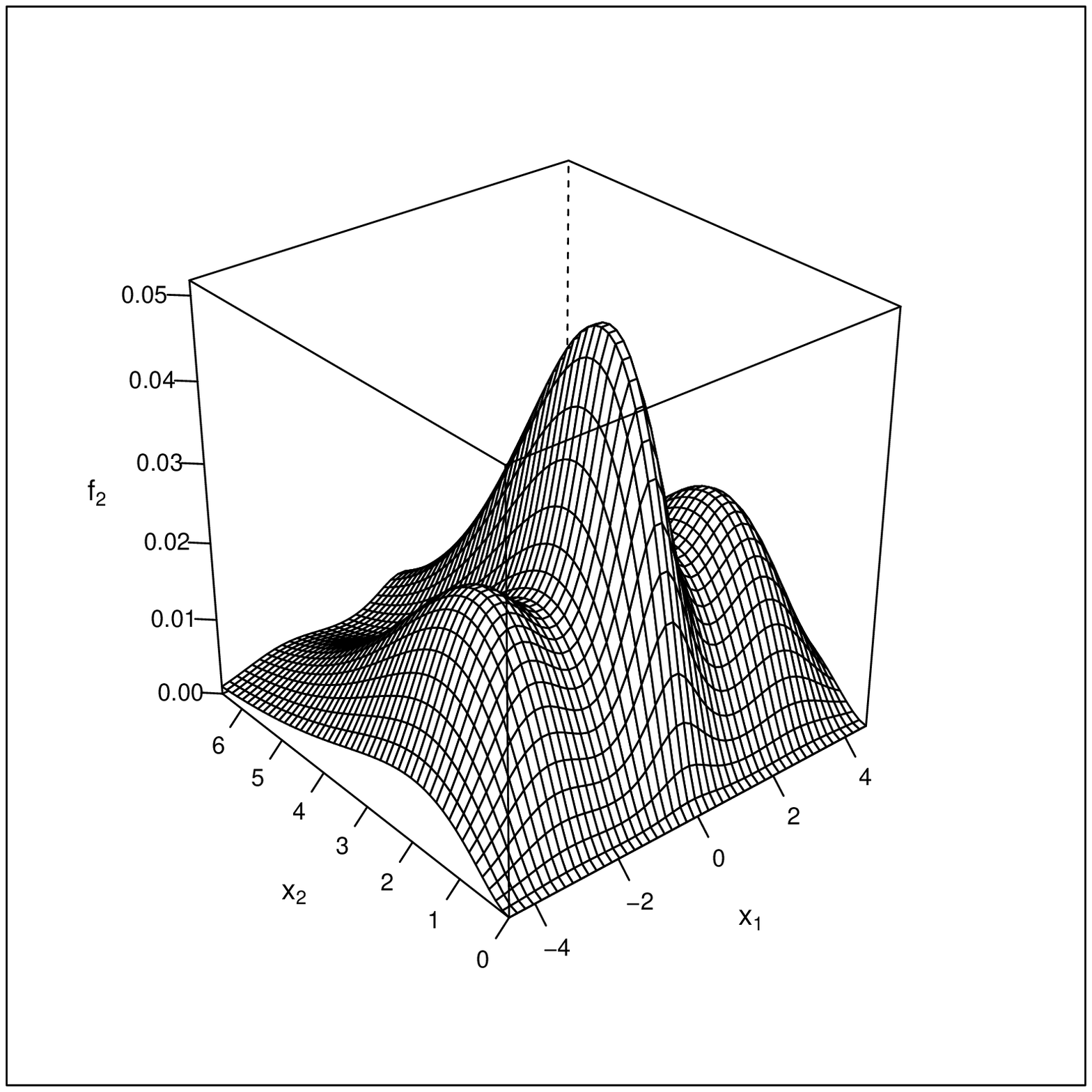}
		\caption{Surface plot of the distribution $f_2$.}\label{f2}
	\end{figure}

	$+ 0.2 \times N (l=3, s=1)$ and $Ga (sh=3,s=1)$, say $f_{2}$ (Fig. \ref{f2}).\\
	$3)$\hspace{.1 in}Product of $0.4 \times N (l=-3, s=3)+ 0.2 \times C (l=0, s=1)$
	
\begin{figure}
		\centering
		\includegraphics[width=1\textwidth]{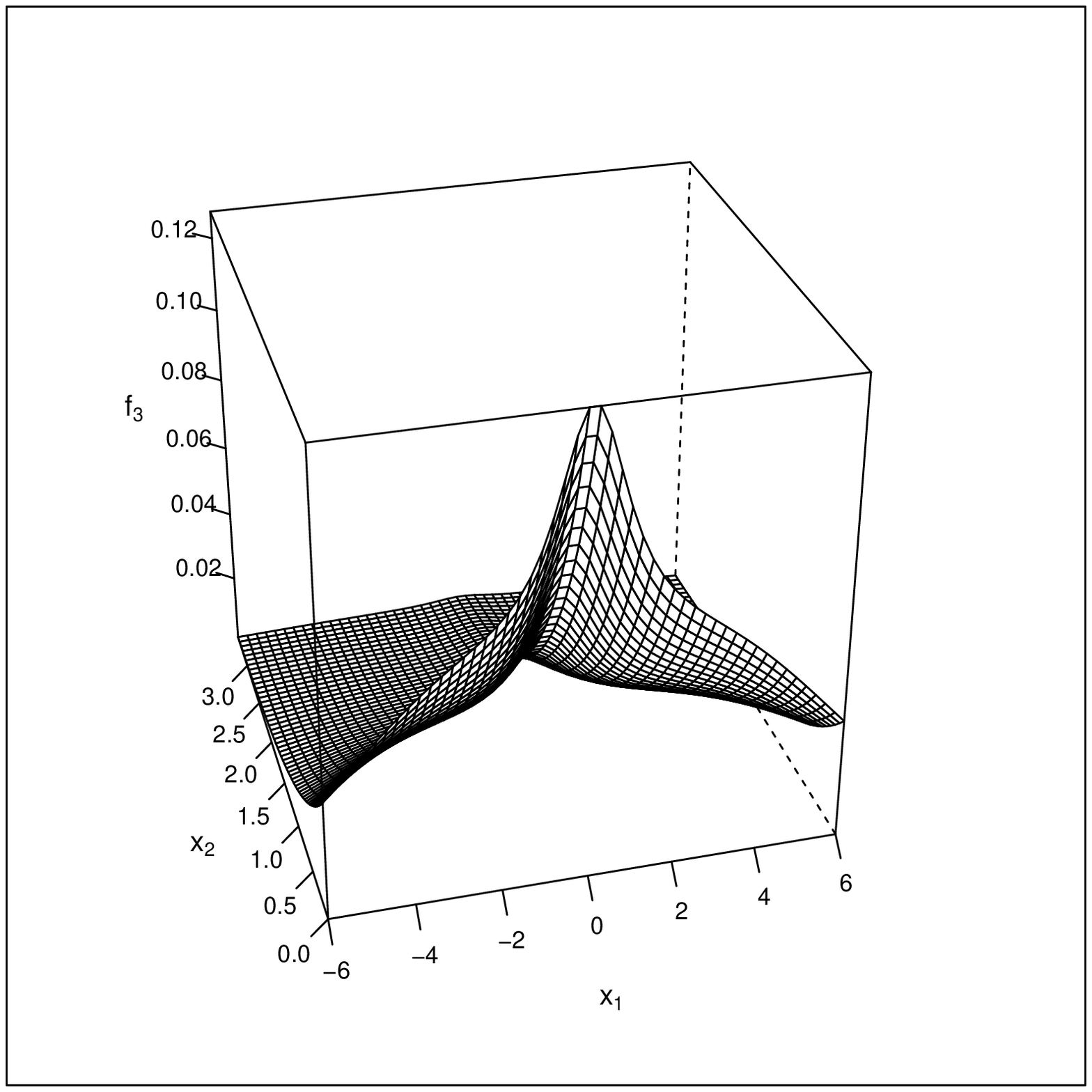}
		\caption{Surface plot of the distribution $f_3$.}\label{f3}
	\end{figure}

	$+ 0.4\times N (l=3, s=3)$ and $Exp (rt=1)$, say $f_{3}$ (Fig. \ref{f3}).\\
	$4)$\hspace{.1 in}Product of $0.5 \times LG (l=-1, s=0.5)+ 0.5 \times LG (l=1.5, s=0.7)$ and
	
\begin{figure}
		\centering
		\includegraphics[width=1\textwidth]{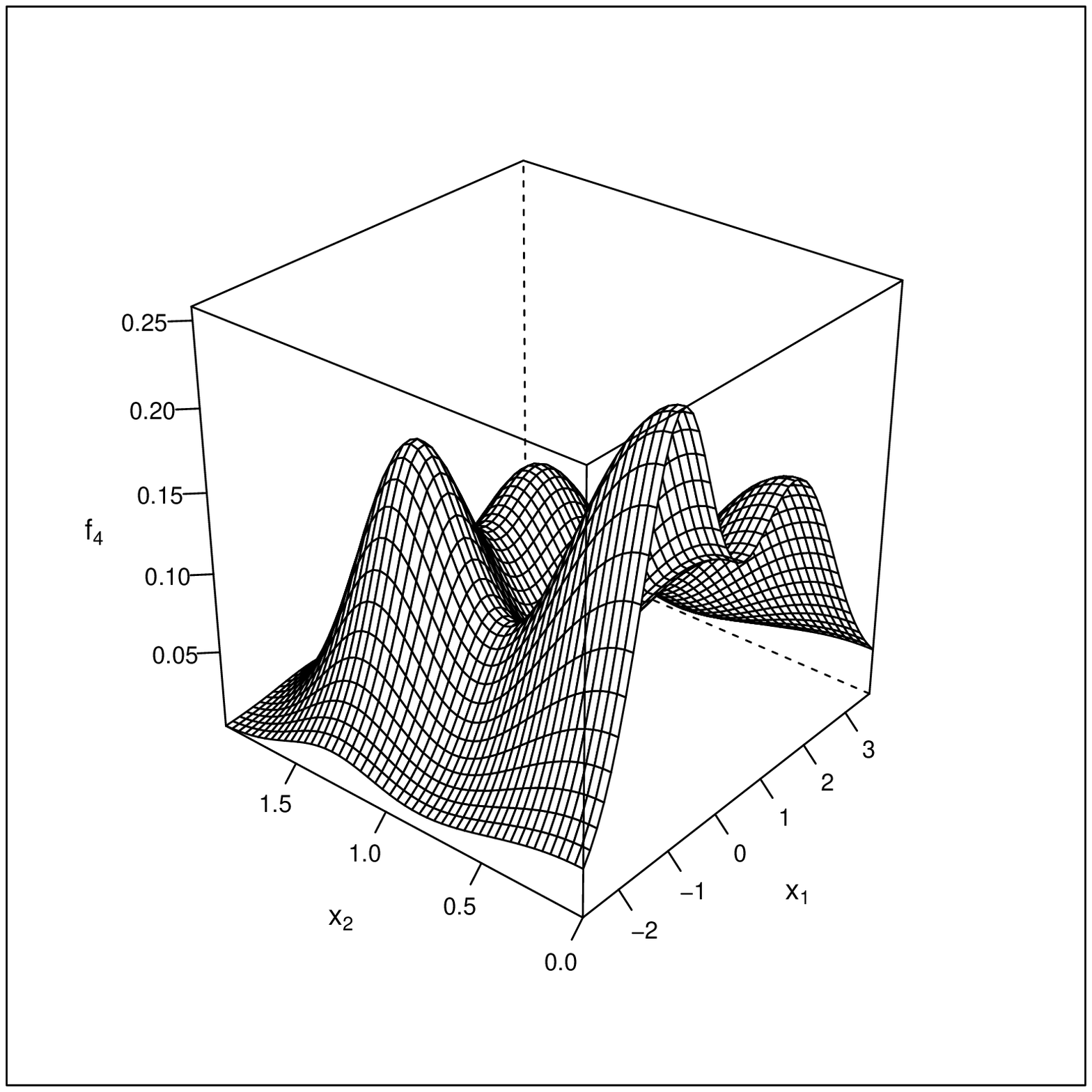}
		\caption{Surface plot of the distribution $f_4$.}\label{f4}
		\end{figure}

	 $0.6 \times TN (l=0, s=0.5)+ 0.4 \times TN (l=1.3, s=0.25)$, say $f_{4}$ (Fig. \ref{f4}).

For comparison among different kernel density estimators, the bandwidths are considered within a range such that $h=b=\sqrt(b_{1})=\sqrt(b_{2})=h_{1}=h_{2}$; where $(h_{1},h_{2})^T$ is the bandwidth vector corresponding to the product of two classical Gaussian $(0,1)$ kernels, which produces the 5--th estimator denoted by $\hat{f_5}$. In each replication the bandwidths are chosen by minimizing the integrated squared error; where the range of integration for each target distribution is specified such that the value of the bivariate density function is ignorable beyond the considered range. The arithmetic mean and the standard deviation of the integrated squared error (ISE) and the bandwidth vector (BW) are reported in Table \ref{t1}.

From the table it is observed, as expected, that with increasing sample size, the mean of ISE and BW of all estimators are decreasing for all target distributions considered. As we can see, $\hat{f}_4$ performs best in all situations and $\hat{f}_3$ does worst among the first four estimators. $\hat{f}_2$ performs second best for the first three target distributions, whereas with distribution $f_4$, $\hat{f}_2$ outperforms $\hat{f}_1$ for $n=100$, but is dominated by $\hat{f}_1$ for $n=200$. This is because $\hat{f}_1$ asymptotically performs better than $\hat{f}_2$ for distribution f$_4$, and the required convergence rate is reached at $n=200$. The first four estimators perform significantly better than $\hat{f}_5$ reflecting the boundary bias problem of classical kernels, except for distribution f$_2$, where $\hat{f}_1$ and $\hat{f}_3$ have mean ISE very close to that of $\hat{f}_5$. In fact, $\hat{f}_3$ is outperformed by $\hat{f}_5$, because distribution f$_2$ has low density near boundary of the non-negative variable, and hence the boundary bias of $\hat{f}_5$ comes out to be close to the biases of $\hat{f}_1$ and $\hat{f}_3$.
\section{Application}\label{5}
Applicability of density estimator $\hat{f}_4$ is demonstrated using two sets of astronomical data. For given data set $\textbf{X}_{1}=(X_{11},X_{12})^T, \ldots,\textbf{X}_{n}=(X_{n1},X_{n2})^T$ of size $n$, the bandwidths are selected by minimizing
$$\int \hat{f}^2(\textbf{x})d\textbf{x}-\frac{2}{n^2}\sum_{i}\hat{f}_{-i}(\textbf{X}_i),$$ where $\hat{f}_{-i}(\textbf{X}_i)$ is an estimate of the target distribution $f$ at the point $\textbf{X}_{i}$, based on the given data set excluding the observation $\textbf{X}_{i}$.

\begin{figure}
		\centering
		\includegraphics[width=1\textwidth]{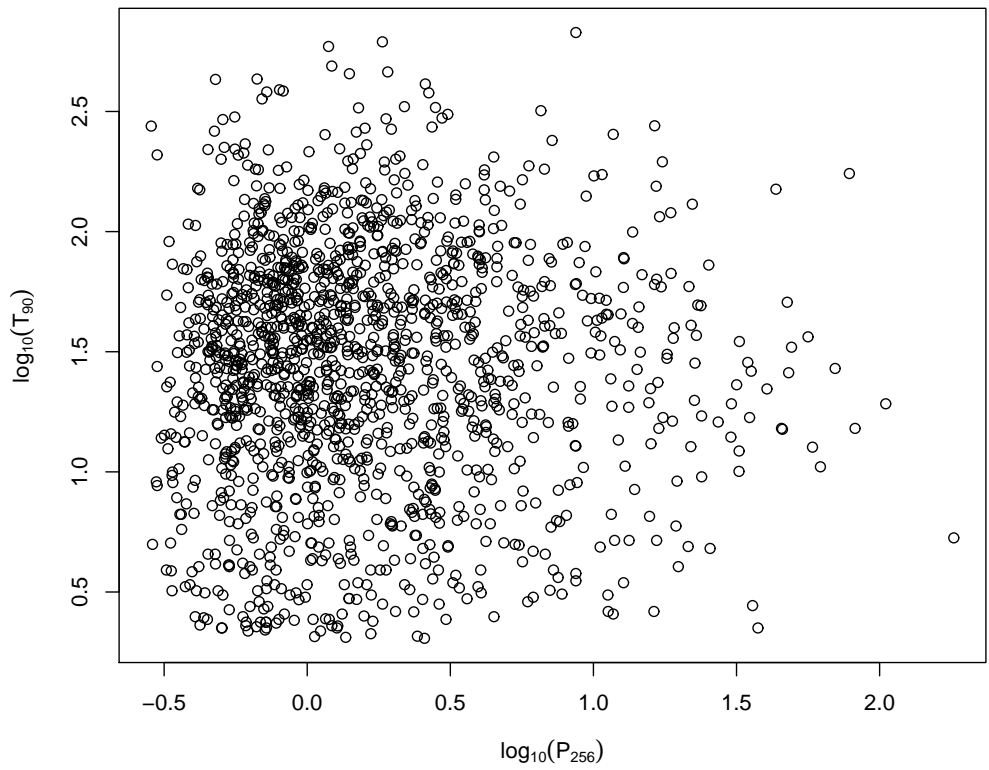}
		\caption{Scatter plot of the GRB data set.}\label{f5}
		\end{figure}

\begin{figure}
		\centering
		\includegraphics[width=1\textwidth]{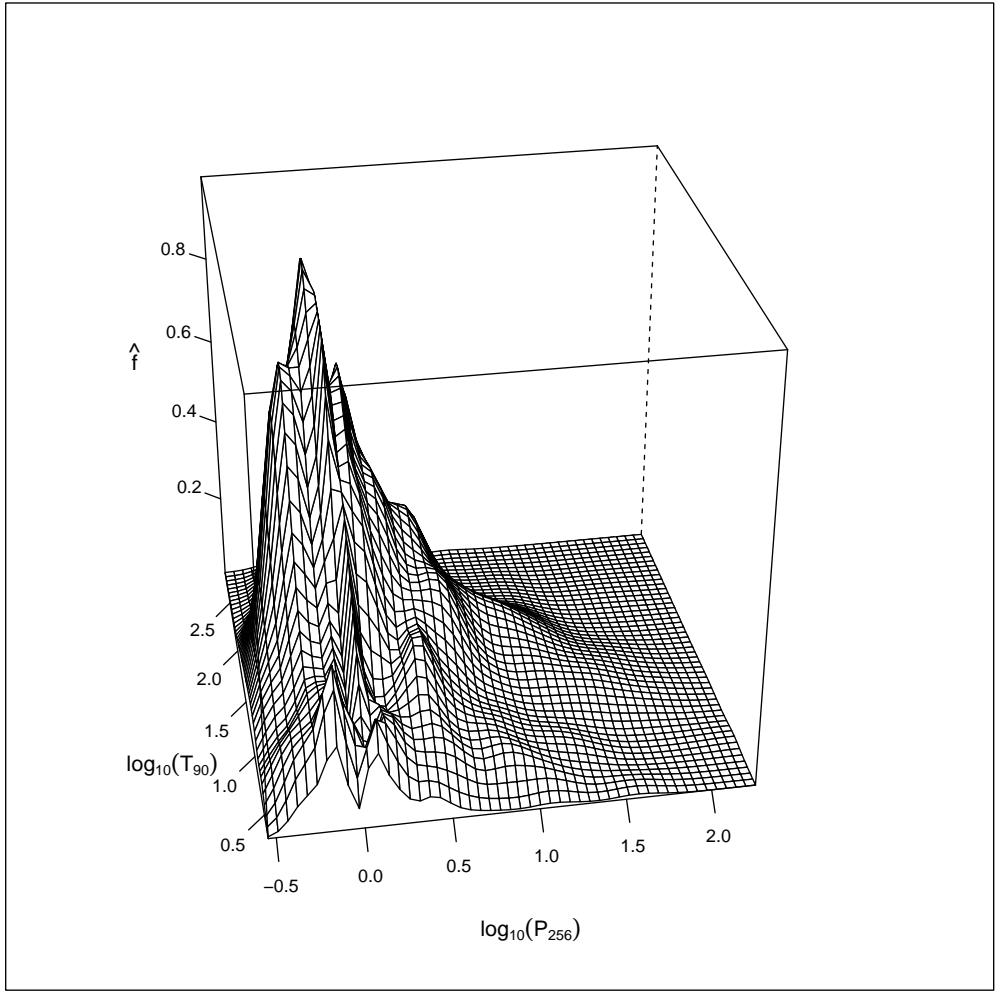}
		\caption{Surface plot of the density of GRB data set, estimated by $\hat{f}_4$ with BW$=(0.013, 0.016)$.}\label{f6}
	\end{figure}

Our first data set consists of information on gamma-ray bursts (GRBs), the brightest explosion in the universe to date since the Big Bang (see, for example, Modak et al. 2018). We collect data from the fourth BATSE Gamma-Ray Burst Catalog (revised) (Paciesas et al., $1999$) for the variable $T_{90}$, a measure of burst duration, is the time in second within which $90\%$ of the flux arrive, and the variable $P_{256}$, peak flux measured in count per square centimeter per second on the $256$ millisecond time scale, on $1496$ long-duration GRBs (i.e. GRBs with $T_{90}>2$ seconds). As usually done in astronomical studies to analyze the data set with huge variation in the values of the variables, we also consider the logarithm transformation of the variables, and study the relation between $log_{10}(T_{90})$ and $log_{10}(P_{256})$ (see, Fig.~\ref{f5}). Fig.~\ref{f6} shows their bivariate density, estimated by $\hat{f}_4$, corresponds to a multimodal distribution.

\begin{figure}
		\centering
		\includegraphics[width=1\textwidth]{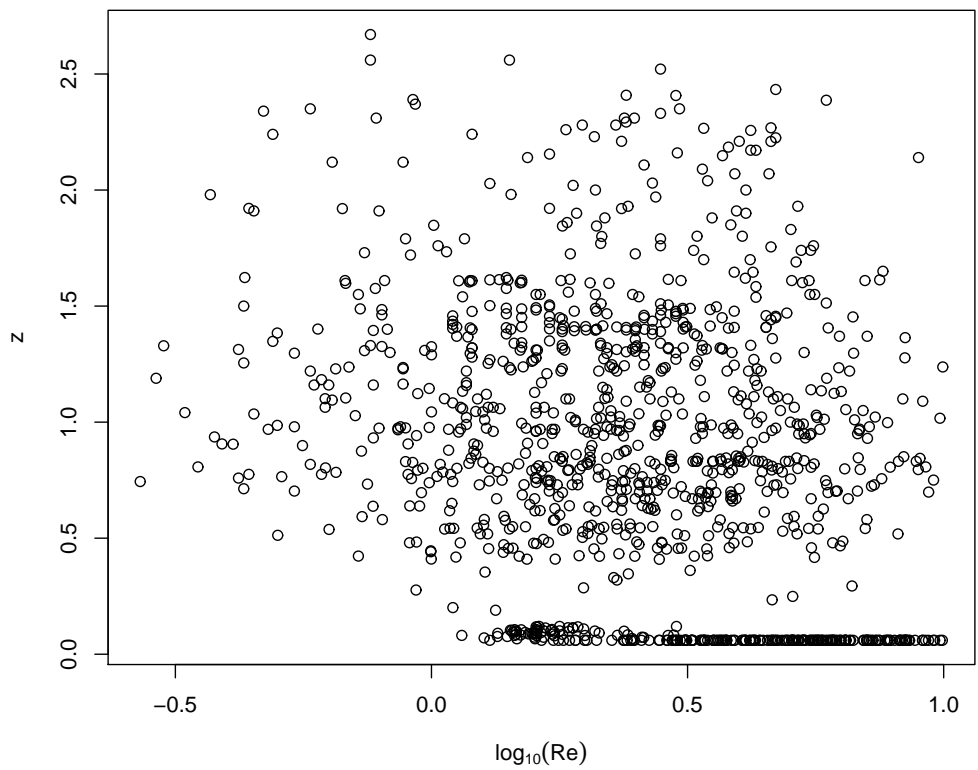}
		\caption{Scatter plot of the ETG data set.}\label{f7}
		\end{figure}

\begin{figure}
		\centering
		\includegraphics[width=1\textwidth]{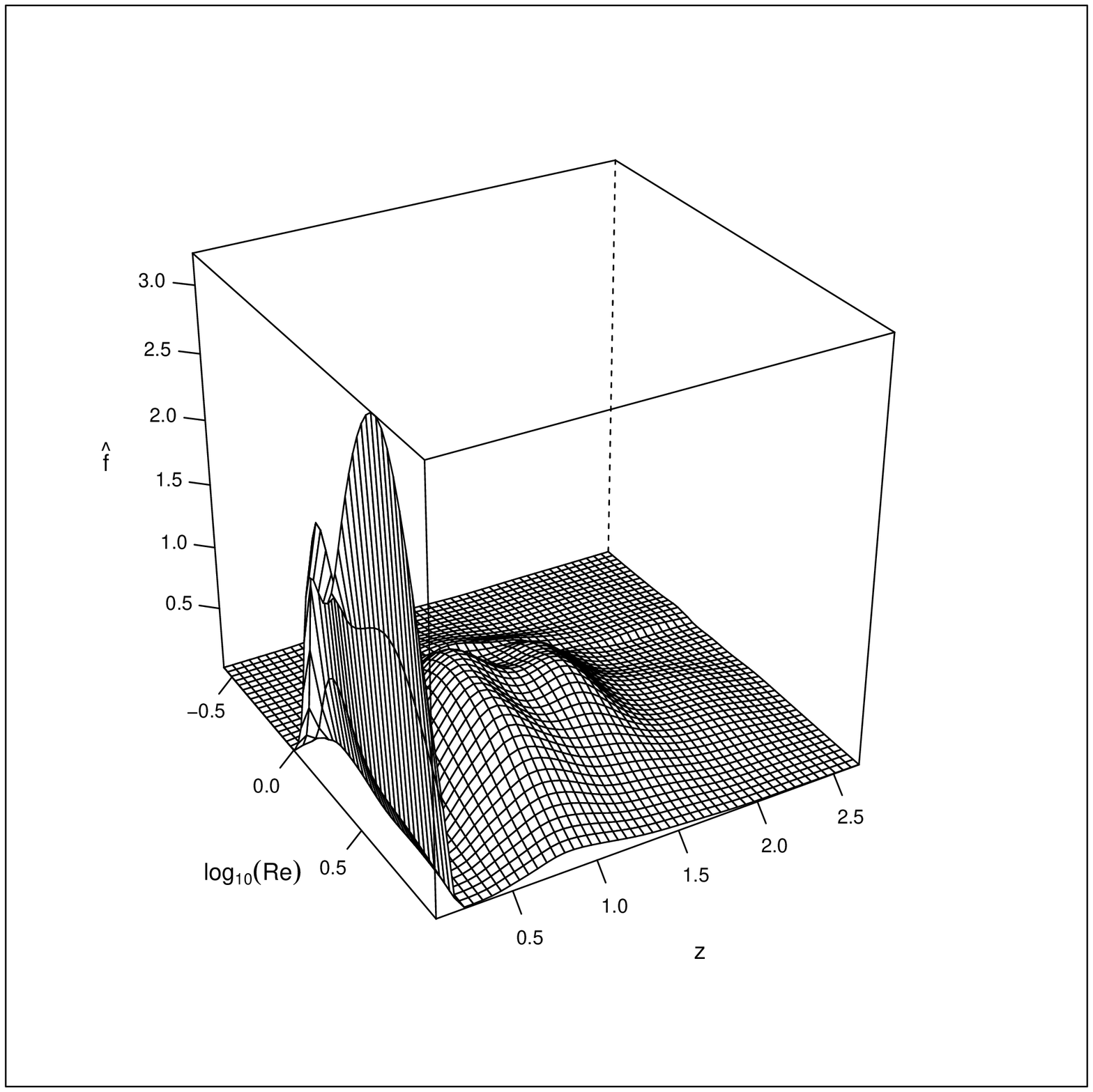}
		\caption{Surface plot of the density of ETG data set, estimated by $\hat{f}_4$ with BW$=( 0.02, 0.02)$.}\label{f8}
		\end{figure}

Next data set contains information on $1131$ early-type galaxies (ETGs) at redshift $(z)$ ranging from $0.06$ to $2.67$ (see, for example, Modak et al. 2017), where the data is collected from F\"{o}rster et al., $2009$; Saracco et al., $2009$; Taylor et al., $2010$; Damjanov et al., $2011$; Papovich et al., $2012$; Chen et al., $2013$; McLure et al., $2013$, and Szomoru et al., $2013$. Fig.~\ref{f7} shows the plot of $log_{10}Re$ ($Re$: effective radius in kiloparsec) versus $z$ for the galaxies, and the corresponding bivariate density estimated by $\hat{f}_4$ is shown in Fig.~\ref{f8}. The latter figure indicates two significantly different patterns in the density function, in which there is a high density function for the nearby ETGs, i.e. galaxies with redshift close to zero, and a low density function for the ETGs in higher redshift region.	
\section{Discussion}\label{6}
We consider the estimation of a bivariate density function with support $\Re\times[0,\infty)$ using the product of classical and associated kernels. We also suggest two new nonparametric density estimators based on $NG$ kernels. The proposed estimators are proved to be free from boundary bias, whereas simulation study shows the estimator $\hat{f}_4$ performs best among its possible competitors. Practical implementation of estimator $\hat{f}_4$ includes two important fields of astronomical study, viz. GRB and ETG. In simulation, the bandwidth vector is selected by optimization method such that the integrated squared error is minimized. With increasing dimension, even if for the bivariate case, this selection procedure can be time consuming. Hence a fast algorithm is needed and bandwidth selection based on the specific kernel estimator is also desired. A search for full bandwidth matrix instead of diagonal matrix can be an open problem for future work (see, for example, Kokonendji and Som\'{e} $2018$).
\clearpage
	\begin{table} \centering
	\tiny
	\caption{Simulation study of the kernel density estimators}
	\label{t1}
	\begin{tabular}{@{\extracolsep{5pt}} ccccccc}
		\\[-1.8ex]\hline
		\hline \\[-1.8ex]
		Estimator & ISE (mean)      & ISE (sd)      & $1^{st}$ BW  & $1^{st}$ BW  & $2^{nd}$ BW  & $2^{nd}$ BW  \\
		& $\times10^6$    & $\times10^6$  &   (mean)     &     (sd)     &     (mean)   & (sd)\\
		\hline \\[-1.8ex]
		&&&n$=100$, distribution$f_{1}$&&&\\
		$\hat{f}_1$ & $2,984$ & $1,360$ & $0.481$ & $0.072$ & $0.358$ & $0.095$ \\
		$\hat{f}_2$ & $2,644$ & $1,435$ & $0.520$ & $0.090$ & $0.576$ & $0.158$ \\
		$\hat{f}_3$ & $3,058$ & $1,414$ & $0.302$ & $0.072$ & $0.208$ & $0.060$ \\
		$\hat{f}_4$ & $2,384$ & $1,282$ & $0.230$ & $0.070$ & $0.390$ & $0.113$ \\
		$\hat{f}_5$ & $4,476$ & $1,620$ & $0.516$ & $0.073$ & $0.485$ & $0.110$ \\
		
		&&&n$=200$, distribution$f_{1}$&&&\\
		$\hat{f}_1$ & $2,024$ & $893$ & $0.418$ & $0.053$ & $0.283$ & $0.071$ \\
		$\hat{f}_2$ & $1,720$ & $880$ & $0.447$ & $0.068$ & $0.483$ & $0.142$ \\
		$\hat{f}_3$ & $2,108$ & $954$ & $0.247$ & $0.050$ & $0.169$ & $0.042$ \\
		$\hat{f}_4$ & $1,608$ & $822$ & $0.184$ & $0.052$ & $0.341$ & $0.097$ \\
		$\hat{f}_5$ & $3,278$ & $1,056$ & $0.457$ & $0.060$ & $0.392$ & $0.084$ \\
		
		&&&n$=100$, distribution$f_{2}$&&&\\
		$\hat{f}_1$ & $1,700$ & $475$ & $1.105$ & $0.463$ & $0.211$ & $0.072$ \\
		$\hat{f}_2$ & $1,607$ & $460$ & $1.035$ & $0.425$ & $0.249$ & $0.074$ \\
		$\hat{f}_3$ & $1,720$ & $566$ & $0.950$ & $0.444$ & $0.116$ & $0.052$ \\
		$\hat{f}_4$ & $1,480$ & $469$ & $0.719$ & $0.348$ & $0.176$ & $0.044$ \\
		$\hat{f}_5$ & $1,716$ & $464$ & $1.015$ & $0.432$ & $0.722$ & $0.148$ \\
		
		&&&n$=200$, distribution$f_{2}$&&&\\
		$\hat{f}_1$ & $1,216$ & $341$ & $0.729$ & $0.272$ & $0.185$ & $0.049$ \\
		$\hat{f}_2$ & $1,137$ & $334$ & $0.683$ & $0.227$ & $0.224$ & $0.050$ \\
		$\hat{f}_3$ & $1,272$ & $361$ & $0.613$ & $0.343$ & $0.099$ & $0.039$ \\
		$\hat{f}_4$ & $1,071$ & $306$ & $0.439$ & $0.232$ & $0.160$ & $0.034$ \\
		$\hat{f}_5$ & $1,242$ & $343$ & $0.694$ & $0.242$ & $0.663$ & $0.104$ \\
		
		&&&n$=100$, distribution$f_{3}$&&&\\
		$\hat{f}_1$ & $2,271$ & $990$ & $1.611$ & $0.442$ & $0.151$ & $0.053$ \\
		$\hat{f}_2$ & $1,985$ & $1,108$ & $1.964$ & $0.549$ & $0.204$ & $0.056$ \\
		$\hat{f}_3$ & $2,303$ & $1,094$ & $1.115$ & $0.428$ & $0.081$ & $0.039$ \\
		$\hat{f}_4$ & $1,701$ & $1,037$ & $1.035$ & $0.396$ & $0.126$ & $0.048$ \\
		$\hat{f}_5$ & $4,160$ & $1,301$ & $1.987$ & $0.521$ & $0.173$ & $0.069$ \\
		
		&&&n$=200$, distribution$f_{3}$&&&\\
		$\hat{f}_1$ & $1,667$ & $669$ & $1.293$ & $0.374$ & $0.128$ & $0.040$ \\
		$\hat{f}_2$ & $1,439$ & $722$ & $1.580$ & $0.487$ & $0.177$ & $0.050$ \\
		$\hat{f}_3$ & $1,663$ & $731$ & $0.881$ & $0.343$ & $0.067$ & $0.030$ \\
		$\hat{f}_4$ & $1,207$ & $653$ & $0.846$ & $0.319$ & $0.107$ & $0.037$ \\
		$\hat{f}_5$ & $3,207$ & $909$ & $1.754$ & $0.445$ & $0.137$ & $0.057$ \\
		&&&n$=100$, distribution$f_{4}$&&&\\
		$\hat{f}_1$ & $10,941$ & $2,417$ & $0.729$ & $0.191$ & $0.071$ & $0.039$ \\
		$\hat{f}_2$ & $10,791$ & $2,347$ & $0.738$ & $0.212$ & $0.151$ & $0.153$ \\
		$\hat{f}_3$ & $11,875$ & $2,791$ & $0.429$ & $0.173$ & $0.045$ & $0.025$ \\
		$\hat{f}_4$ & $10,574$ & $2,420$ & $0.333$ & $0.150$ & $0.085$ & $0.055$ \\
		$\hat{f}_5$ & $12,308$ & $3,115$ & $0.749$ & $0.179$ & $0.147$ & $0.039$ \\
		
		&&&n$=200$, distribution$f_{4}$&&&\\
		$\hat{f}_1$ & $8,183$ & $1,863$ & $0.593$ & $0.120$ & $0.048$ & $0.024$ \\
		$\hat{f}_2$ & $8,339$ & $1,889$ & $0.619$ & $0.132$ & $0.075$ & $0.081$ \\
		$\hat{f}_3$ & $8,860$ & $2,185$ & $0.295$ & $0.101$ & $0.034$ & $0.015$ \\
		$\hat{f}_4$ & $8,101$ & $1,867$ & $0.248$ & $0.091$ & $0.056$ & $0.029$ \\
		$\hat{f}_5$ & $9,322$ & $2,262$ & $0.607$ & $0.113$ & $0.124$ & $0.024$ \\
		\hline \\[-1.8ex]
	\end{tabular}
\end{table}
\clearpage
\clearpage
\clearpage
\clearpage
\clearpage
\clearpage
\clearpage
\clearpage
	\clearpage
\section*{Appendix}
\subsection*{A.1}
\begin{align*}
\lefteqn{\{E_{\xi_{x_1},\xi_{x_2}} (\xi_{x_1}-x_1)\} f^1(\textbf{x})
	=f^1(\textbf{x})\{ E_{\xi_{x_1}|\xi_{x_2},\xi_{x_2}} (\xi_{x_1}-x_1)\}}\\
&=f^1(\textbf{x})\Bigg[\int_{\xi_{x_1}} \int_{\xi_{x_2}}(\xi_{x_1}-x_1) N(\xi_{x_1}|\mu,(\lambda\xi_2)^{-1}) Ga(\xi_{x_2}|\alpha,\beta) d\xi_{x_1} d\xi_{x_2}\Bigg]\\
&=f^1(\textbf{x})\Bigg[ \int_{\xi_{x_2}}\Bigg\{\int_{\xi_{x_1}} (\xi_{x_1}-x_1) N(\xi_{x_1}|\mu,(\lambda\xi_2)^{-1})d\xi_{x_1}\Bigg\} Ga(\xi_{x_2}|\alpha,\beta)  d\xi_{x_2}\Bigg]\\
&=f^1(\textbf{x})\Bigg[ \int_{\xi_{x_2}}\{E_{\xi_{x_1}|\xi_{x_2}} (\xi_{x_1}-E_{\xi_{x_1}|\xi_{x_2}}(\xi_{x_1}))\} Ga(\xi_{x_2}|\alpha,\beta)  d\xi_{x_2}\Bigg]\\
&=0,   \quad \text{since} \quad {\xi_{x_1}|\xi_{x_2}}{\hspace{.05in}\text{follows}\hspace{.05in}}  N(\xi_{x_1}|\mu,(\lambda\xi_2)^{-1}).
\end{align*}
Similarly,
\begin{equation*}
\{E_{\xi_{x_1},\xi_{x_2}} (\xi_{x_2}-x_2)\} f^2(\textbf{x})=2b_2f^2(\textbf{x}).
\end{equation*}
Again,
\begin{align*}
\lefteqn{\frac{1}{2}E\{(\xi_{x_1}-x_1)(\xi_{x_2}-x_2)\} f^{12}(\textbf{x})}\\
&=\frac{1}{2}f^{12}(\textbf{x})\Bigg[\int_{\xi_{x_2}}(\xi_{x_2}-x_2)\Bigg\{\int_{\xi_{x_1}} (\xi_{x_1}-x_1) N(\xi_{x_1}|\mu,(\lambda\xi_2)^{-1})d\xi_{x_1}\Bigg\}Ga(\xi_{x_2}|\alpha,\beta)  d\xi_{x_2}\Bigg]\\
&=0.
\end{align*}
Similarly,
\begin{align*}
\frac{1}{2}E\{(\xi_{x_1}-x_1)(\xi_{x_2}-x_2)\} f^{21}(\textbf{x})&=0.
\end{align*}
Now,
\begin{align*}
\lefteqn{\frac{1}{2}E\{(\xi_{x_1}-x_1)^2\} f^{11}(\textbf{x})}\\
&= \frac{1}{2}f^{11}(\textbf{x})\Bigg[\int_{\xi_{x_2}}\Bigg\{\int_{\xi_{x_1}} (\xi_{x_1}-x_1)^2 N(\xi_{x_1}|\mu,(\lambda\xi_2)^{-1})d\xi_{x_1}\Bigg\} Ga(\xi_{x_2}|\alpha,\beta)  d\xi_{x_2}\Bigg]\\
&= \frac{1}{2}f^{11}(\textbf{x})\Bigg[\int_{\xi_{x_2}}\{Var_{\xi_{x_1}|\xi_{x_2}}(\xi_{x_1})\}Ga(\xi_{x_2}|\alpha,\beta)  d\xi_{x_2}\Bigg]\\
&= \frac{1}{2}f^{11}(\textbf{x}) \lambda^{-1}\Bigg\{\int_{\xi_{x_2}}\xi_{x_2}^{-1} Ga(\xi_{x_2}|\alpha,\beta)  d\xi_{x_2}\Bigg\},\hspace{.05in}\text{as} \hspace{.05in} Var_{\xi_{x_1}|\xi_{x_2}}(\xi_{x_1})=(\lambda \xi_{x_2})^{-1}\\
&=\frac{1}{2}f^{11}(\textbf{x})\lambda^{-1}(\alpha-1)^{-1}\beta
= \frac{1}{2}|x_1|f^{11}(\textbf{x}).
\end{align*}
Similarly,
\begin{align*}
\lefteqn{\frac{1}{2}E\{(\xi_{x_2}-x_2)^2\} f^{22}(\textbf{x})
	=\frac{1}{2}f^{22}(\textbf{x})\{Var(\xi_{x_2})+(2b_2)^2\}}\\
&= \frac{1}{2}f^{22}(\textbf{x})(x_2b_2+6b_2^2),\hspace{.05in}\text{since}\hspace{.05in} Var(\xi_{x_2})=x_2b_2+2b_2^2.
\end{align*}
\subsection*{A.2}
\begin{align*}
Var\{\hat{f}(\textbf{x})\}&=n^{-1} Var\{ K_{\Theta}(\textbf{X}_{i})\}\\
&=n^{-1}E\{  K_{\Theta}(\textbf{X}_{i})\}^2+O(n^{-1}).\\
\end{align*}
Now,
\begin{align*}
E_{\xi_{x_1},\xi_{x_2}}\{K_{\Theta}(\textbf{X}_{i})\}^2&= \int_{\xi_{x_2}}\Bigg[\int_{\xi_{x_1}} \{N(\xi_{x_1}|\mu,(\lambda\xi_2)^{-1}\}^2 d\xi_{x_1}\Bigg]  \{Ga(\xi_{x_2}|\alpha,\beta)\}^2  d\xi_{x_2}\\
&=\frac{\sqrt{\lambda}\sqrt{\eta_{x_2}}}{2\sqrt{\pi}} N(\eta_{x_1}|\mu,(2\lambda\eta_{x_2})^{-1}) \{Ga(\xi_{x_2}|\alpha,\beta)\}^2  d\xi_{x_2}\\
&=B_{b_1,b_2}(x_1,x_2) N(\eta_{x_1}|\mu,(2\lambda\eta_{x_2})^{-1})  Ga(\eta_{x_2}|2\alpha-1/2,2\beta),
\end{align*}
where $B_{b_1,b_2}(x_1,x_2)=\frac{\sqrt{\lambda}\sqrt{\Gamma(2\alpha-1/2)}}{\sqrt{\pi}2^{2\alpha+1/2}\Gamma(\alpha)^2} \beta^{1/2}$. Then, writing the gamma function in terms of $R(z)=\sqrt{2\pi}e^{-z}z^{z+1/2}/\Gamma(z+1)$ for $z\geq0$ we obtain,
\begin{align*}
\frac{\sqrt{\Gamma(2\alpha-1/2)}}{\Gamma(\alpha)^2}&=\frac{1}{\sqrt{\pi e}}2^{2\alpha-3/2}\Bigg\{1+\frac{1}{4(\alpha-1)}\Bigg\}^{2\alpha-1}\frac{R^2(\alpha-1)}{R(2\alpha-3/2)},
\end{align*}
which implies (Brown and Chen, $1999$)  $B_{b_1,b_2}(x_1,x_2)\leq \frac{1}{4\pi\sqrt{ e}}\sqrt{\lambda\beta}$ for $\alpha\rightarrow\infty$. Therefore, $ B_{b_1,b_2}(x_1,x_2)$ is approximated as
\[\begin{cases}
\frac{1}{4\pi\sqrt{ e}}b_1^{-1/2}  b_2^{-1/2}|x_1|^{-1/2} x_2^{-1/2}        & \quad \text{if } x_1/b_1\rightarrow\infty \quad \text{and} \quad x_2/b_2\rightarrow\infty,\\
\frac{\Gamma(2\kappa_2+7/2)}{\sqrt{\pi}2^{2\kappa_2+9/2}\sqrt{(\kappa_1+1)(\kappa_2+1)}\Gamma^2(\kappa_2+2)} b_1^{-1} b_2^{-1}    & \quad \text{if }  x_1/b_1\rightarrow\kappa_1\quad \text{and} \quad x_2/b_2\rightarrow\kappa_2,\\
\frac{\Gamma(2\kappa_2+7/2)}{\sqrt{\pi}2^{2\kappa_2+9/2}\sqrt{(\kappa_2+1)}\Gamma^2(\kappa_2+2)} b_1^{-1/2} b_2^{-1}|x_1|^{-1/2}    & \quad \text{if }  x_1/b_1\rightarrow\infty\quad \text{and} \quad x_2/b_2\rightarrow\kappa_2,\\
\frac{1}{4\pi\sqrt{ e}\sqrt{\kappa_1+1}}b_1^{-1}  b_2^{-1/2} x_2^{-1/2}        & \quad \text{if } x_1/b_1\rightarrow\kappa_1 \quad \text{and} \quad x_2/b_2\rightarrow\infty,\\
\end{cases}\]
for non-negative constants $\kappa_1,\kappa_2$.
\subsection*{A.3}
Now, for $\delta_1=b_1^{1-\epsilon_1}$ with $0<\epsilon_1<1$ and $\delta_2=b_2^{1-\epsilon_2}$ with $0<\epsilon_2<1,$
\begin{align*}
\lefteqn{\int_{-\infty}^{\infty} \int_{0}^{\infty} Var\{\hat{f_3}(\textbf{x})\} dx_1 dx_2=2\int_{0}^{\infty} \int_{0}^{\infty} Var\{\hat{f_3}(\textbf{x})\} dx_1 dx_2}\\
&=2\Bigg[\int_{0}^{\delta_1} \int_{0}^{\delta_2} Var\{\hat{f_3}(\textbf{x})\} dx_1 dx_2+\int_{0}^{\delta_1} \int_{\delta_2}^{\infty} Var\{\hat{f_3}(\textbf{x})\} dx_1 dx_2\\
&\hspace{.165in}+\int_{\delta_1}^{\infty} \int_{0}^{\delta_2} Var\{\hat{f_3}(\textbf{x})\} dx_1 dx_2+\int_{\delta_1}^{\infty} \int_{\delta_2}^{\infty} Var\{\hat{f_3}(\textbf{x})\} dx_1 dx_2\Bigg]\\
&=2\Bigg[\frac{1}{4\pi\sqrt{ e}\sqrt{\kappa_1+1}}n^{-1}b_1^{-1}  b_2^{-1/2} \int_{0}^{\delta_1} \int_{\delta_2}^{\infty}x_2^{-1/2} f(\textbf{x}) dx_1 dx_2\\
&\hspace{.165in}+\frac{\Gamma(2\kappa_2+7/2)}{\sqrt{\pi}2^{2\kappa_2+9/2}\sqrt{(\kappa_2+1)}\Gamma^2(\kappa_2+2)}n^{-1} b_1^{-1/2} b_2^{-1}\int_{\delta_1}^{\infty} \int_{0}^{\delta_2} |x_1|^{-1/2}  f(\textbf{x})  dx_1 dx_2\\
&\hspace{.165in}+\frac{1}{4\pi\sqrt{ e}}n^{-1}b_1^{-1/2}  b_2^{-1/2}\int_{\delta_1}^{\infty} \int_{\delta_2}^{\infty}|x_1|^{-1/2} x_2^{-1/2} f(\textbf{x}) dx_1 dx_2+O(n^{-1}b_{1}^{-\epsilon_1}b_{2}^{-\epsilon_2})\Bigg]\\
&=2\Bigg[\frac{1}{4\pi\sqrt{ e}}n^{-1}b_1^{-1/2}  b_2^{-1/2}\int_{0}^{\infty} \int_{0}^{\infty}|x_1|^{-1/2} x_2^{-1/2} f(\textbf{x}) dx_1 dx_2+o(n^{-1}b_{1}^{-1/2}b_{2}^{-1/2})\Bigg]\\
&=\frac{1}{4\pi\sqrt{ e}}n^{-1}b_1^{-1/2}  b_2^{-1/2}\int_{-\infty}^{\infty} \int_{0}^{\infty}|x_1|^{-1/2} x_2^{-1/2} f(\textbf{x}) dx_1 dx_2+o(n^{-1}b_{1}^{-1/2}b_{2}^{-1/2})
\end{align*}
with $\int_{-\infty}^{\infty}  \int_{0}^{\infty}|x_1|^{-1/2} x^{-1/2}_2 f(\textbf{x}) dx_1 dx_2$ finite.\\

\end{document}